\documentclass[runningheads]{llncs}
\usepackage{enumitem}   
\usepackage{graphicx}
\usepackage{subcaption}
\usepackage[T1]{fontenc}
\usepackage{graphicx}
\begin{document}
\title{CovHuSeg: An Enhanced Approach for Kidney Pathology Segmentation}
%
%
\author{Huy Trinh\inst{1,2\dagger} \and
Khang Tran\inst{1,2\dagger} \and
Nam Nguyen\inst{2,3}\and
Tri Cao\inst{1,2}\and
Binh Nguyen\inst{1,2}\thanks{Correspondong author: Binh T. Nguyen (ngtbinh@hcmus.edu.vn), $\dagger$ These authors contributed equally to this work.}}
\authorrunning{Huy et al.}
%
\newcommand{\RomanNumeralCaps}[1]
{\MakeUppercase{\romannumeral #1}}
\institute{
University of Science, Ho Chi Minh City, Vietnam 
\and
Vietnam National University Ho Chi Minh City, Vietnam
\and
VNU-HCM High School for the Gifted, Vietnam 
}
\maketitle              
%
\begin{abstract}

Segmentation has long been essential in computer vision due to its numerous real-world applications. However, most traditional deep learning and machine learning models need help to capture geometric features such as size and convexity of the segmentation targets, resulting in suboptimal outcomes. To resolve this problem, we propose using a CovHuSeg algorithm to solve the problem of kidney glomeruli segmentation. This simple post-processing method is specified to adapt to the segmentation of ball-shaped anomalies, including the glomerulus. Unlike other post-processing methods, the CovHuSeg algorithm assures that the outcome mask does not have holes in it or comes in unusual shapes that are impossible to be the shape of a glomerulus. We illustrate the effectiveness of our method by experimenting with multiple deep-learning models in the context of segmentation on kidney pathology images. The results show that all models have increased accuracy when using the CovHuSeg algorithm.

\end{abstract}

\section{Introduction}
To obtain great accuracy in diagnosis and quantitative analysis in renal pathology, the task of kidney glomeruli segmentation has received many considerations. Recently, deep learning techniques have become essential in this field, helping to enable studies on large-scale population-based  \cite{bueno2020glomerulosclerosis,gadermayr2017cnn,ginley2019computational,govind2018glomerular,kannan2019segmentation,nguyen2023joint,nguyen2023manet,mh2024lvm},  as well as alleviating the clinical workload for pathologists. Numerous traditional, feature-based image processing methods have also been developed for glomeruli segmentation. However, these approaches rely heavily on the work of feature engineering and ``hand-crafted'' features. Some particularly well-known techniques are edge detection \cite{ma2009glomerulus}, Watershed Segmentation \cite{levner2007classification,angulo2007stochastic}, Morphological Operations \cite{comer1999morphological}, Thresholding \cite{kohler1981segmentation} and Region Growing \cite{pohle2001segmentation}.

\begin{figure}[ht]
    \centering
    \includegraphics[width=\textwidth]{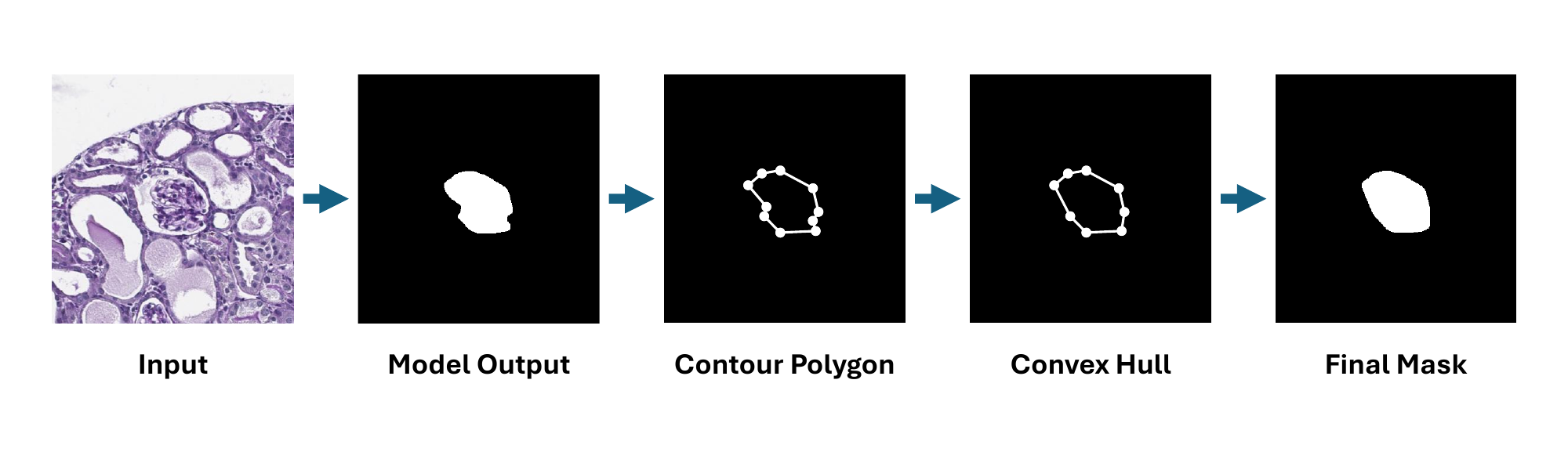}
    \caption{The proposed CovHuSeg algorithm for the kidney glomeruli segmentation.}
    \label{fig:image2}
\end{figure}

Recently, methods based on deep neural networks (CNNs) have proven exceptional performance in glomeruli segmentation by leveraging ``data-driven'' features \cite{kannan2019segmentation,silva2022boundary}. 
However, these models often need to capture geometric features such as convexity or size fully. 
Recognizing this room for improvement, Haichun Yang et al. proposed an anchor-free detection method called CircleNet \cite{yang2020circlenet}, which is optimized for identifying glomeruli in kidney pathology using a bounding circle representation. This approach reduces degrees of freedom and provides rotational invariance. Through various experiments, CircleNet \cite{yang2020circlenet} has shown great detection accuracy and outperforms benchmark models in multiple metrics. Nevertheless, the idea of CircleNet \cite{yang2020circlenet} is only for a detection task. 
When it comes to the problem of segmentation, there are several disadvantages, one of which is that the glomeruli's shape is not always a perfect circle, as seen in the input in Figure \ref{fig:image2}. It could lead to non-optimized accuracy when the idea is used for segmenting.
To tackle this challenge, we have developed and integrated a novel technique called CovHuSeg, which uses convex hulls as a post-processing procedure. Our method ensures that the segmentation mask maintains a convex shape, resulting in more accurate outcomes than the circular shape produced by CircleNet. By applying the CovHuSeg algorithm, we aim to redefine the boundaries of accuracy and efficiency in the segmentation of glomeruli in kidney pathology, particularly in scenarios with limited data.
 
To show the effectiveness of our method, we have conducted multiple experiments in the context of kidney pathology image segmentation using different deep learning models such as UNet \cite{ronneberger2015u}, UNet++ \cite{zhou2018unet++}, UNet3+ \cite{huang2020unet}, TransUNet \cite{chen2024transunet} and different metrics. Furthermore, we also experiment in the scenario where there is a lack of accessible data by training with only 50\% of subjects, 50\% samples per subject, both 50\% of subjects, 50\% sample per subject, and 25\% samples of all subjects in the training dataset. The final results indicate that all models have increased accuracy in all metrics when CovHuSeg post-processing is used, especially in cases with limited data.

The paper is structured comprehensively: Section 1 provides a concise introduction to kidney glomeruli segmentation. Section 2 delves into previous works related to the subject, including state-of-the-art deep-learning models and techniques for adapting to biomedical datasets. In Section 3, we present our algorithm for the main problem. Section 4 is dedicated to illustrating all experimental results and performance comparisons. The paper concludes with our findings and suggestions for further research.

\section{Related Works}
The development of deep learning in recent years has created many breakthroughs in a wide range of fields. In economics, \cite{raghavan2019fraud} has shown the use of deep learning for fraud detection. In cybersecurity, phishing detection by deep learning was studied in \cite{li2024knowphish,cao2024phishagent}. In healthcare and biomedical, deep learning has also been used for drug discovery \cite{gawehn2016deep}, disease detection \cite{koppu2020deep}, and genomics analysis \cite{zou2019primer}. Additionally, one domain where deep learning is particularly utilized is computer vision. Many studies have been conducted in this domain for tasks such as image classification \cite{affonso2017deep}, image segmentation \cite{roth2018deep}, anomaly detection \cite{cao2023anomaly,cao2024anomaly} and facial recognition \cite{singh2020facial} and have shown some great results compared to traditional machine learning methods.

Simultaneous to developing deep learning, many different deep learning models are studied. In 2015, \cite{ronneberger2015u} proposed an architecture named UNet, consisting of a contracting path to capture context and a symmetric expanding path that enables precise localization. This model can be trained with very limited data while still being able to outperform the current time state of the art. Three years later, by adding a series of nested, dense skip pathways to connect the encoder and decoder sub-networks, \cite{zhou2018unet++} has created UNet++. This improvement helped to reduce the semantic
gap between the feature maps of the encoder and decoder sub-networks, which would allow the optimizer to deal with an easier learning task. 

In 2020, another version of UNet named UNet3+ has been proposed in \cite{huang2020unet}. The authors have taken advantage of full-scale skip connections and deep supervision in this version. These changes help to incorporate low-level details with high-level semantics from feature maps in different scales. Another well-known segmentation model is DeepLabv3+ \cite{chen2018encoder}, which is an extension of the conventional DeepLabv3 \cite{chen2017rethinkingatrousconvolutionsemantic} with an additional decoder module to refine the segmentation results, especially along object boundaries. Additionally, many other deep learning models for segmentation were proposed by applying the concept of Transformer \cite{vaswani2017attention}. One of which is MIST, studied in \cite{rahman2024mist}; this model has two parts: a pre-trained multi-axis vision transformer (MaxViT) is used as an encoder, and the encoded feature representation is passed through the CAM decoder for segmenting the images. A few other well-known Transformer models will include TransUNet \cite{chen2024transunet} and UNETR \cite{hatamizadeh2022unetr}, which have also shown great results in terms of accuracy and robustness compared to traditional models.

Although the models mentioned above are widely used and can work with different types of data, certain biomedical datasets, like kidney pathology images, usually need to be optimized due to the complexity and similarity between different images. Many studies have been conducted to resolve this problem. In \cite{yang2020circlenet}, a simple anchor-free detection method with circle representation is proposed for detecting the ball-shaped glomerulus. Similarly, in \cite{chen2021ellipsenet}, the authors consider the bounding box to be ellipse-shaped instead of a circle. Their experiments also show that their proposed method is better than the current state of the art.

One approach to improve deep learning models' accuracy for biomedical datasets that have been studied is to utilize geometric features such as the convexity of the shape of anomalies. This has been witnessed in \cite{10.1007/978-3-031-15037-1_18}, where a model that includes a convex hull filter is applied to segment brain tumors and has been shown to have better accuracy than previous approaches. Additionally, the study in  \cite{article} has proposed a lung contour repair algorithm based on the improved convex hull method to solve the contour loss caused by solid nodules and other lesions. Their technique has resulted in improved accuracy and robustness.

\section{Methodology}
\subsection{Problem Formulation}
Image segmentation \cite{minaee2021image} is a computer vision method that divides a digital image into distinct pixel groups to inform other related tasks, such as object detection. By breaking down an image's complex visual data into clearly defined segments, it enables faster and more advanced image processing.

\begin{figure}[ht]
    \centering
    \includegraphics[width=\textwidth]{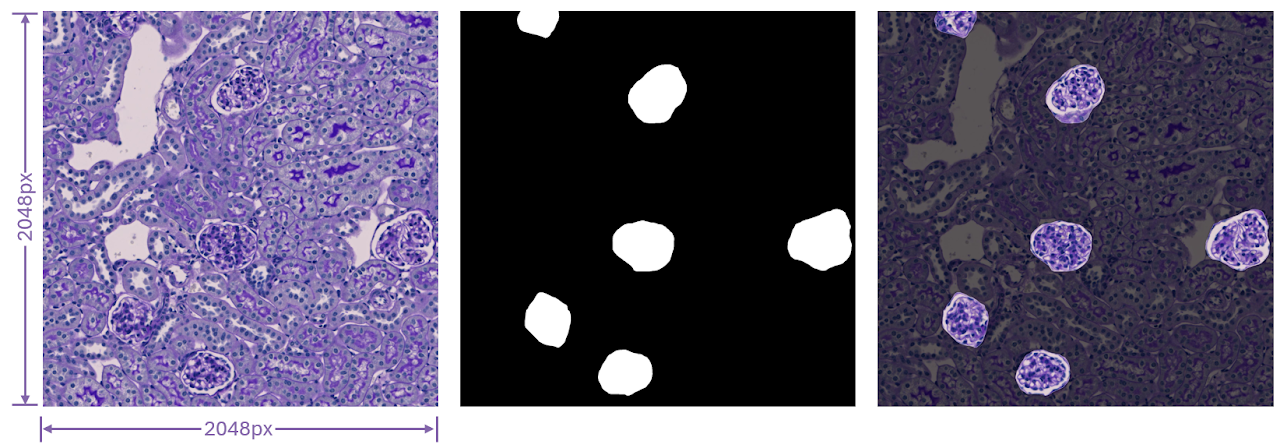}
    \caption{Segmentation Task}
    \label{fig:image3}
\end{figure}

Although numerous studies have focused on the task of segmentation and various segmentation models have been developed, most of these models struggle with biomedical images due to their complexity and similarity. In this work, our task would be to segment glomeruli at a pixel level, adapting to different chronic kidney disease models and tissue conditions. This procedure is shown in Figure \ref{fig:image3}. The effectiveness of baseline models tends to drop significantly when doing this task due to the large number of variations in glomeruli size, shape, and structural integrity, which depend on disease states or preparation techniques. Our objective is to enhance the accuracy of these baseline models by employing our proposed post-processing method: the CovHuSeg algorithm.
\subsection{Convex Hull}
The convex hull of a set of points in an Euclidean space is the smallest convex polygon that encloses all of the points. In two dimensions (2D), the convex hull is a convex polygon, and in three dimensions (3D), it is a convex polyhedron.\\
There are many applications of convex hull in Machine Learning and AI tasks. In Robotics and Path planning, a convex hull helps define the spatial extent of obstacles or objects, aiding in navigation and collision avoidance.  In video processing, convex hulls can be applied to track moving objects by encapsulating their contours and maintaining shape consistency.\\
In this work, we derive an application of the convex hull in biomedical image segmentation. We aim to capture the convex feature of the shape of targeted anomalies and thus improve the overall accuracy of the outcome masks.
\begin{figure}[ht]
    \centering
    \includegraphics[width=\textwidth]{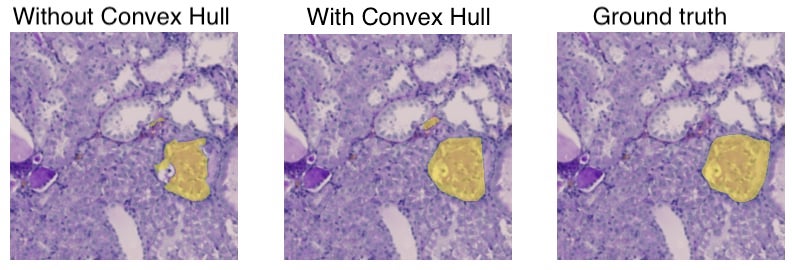}
    \caption{An example of using the CovHuSeg algorithm for the problem of kidney glomeruli segmentation.}
    \label{fig:image1}
\end{figure}
\\
Figure \ref{fig:image1} shows the process of using the CovHuSeg Algorithm to segment an anomaly. It can be seen that without the CovHuSeg, the outcome mask has many holes inside it. Using our method, the returned mask is in a convex shape and is much closer to the ground truth. This is an example of the effectiveness of our algorithm as it helps to improve the accuracy of the segmentation.
\subsection{CovHuSeg Algorithm}
In general, the CovHuSeg Algorithm can be split into four main steps. First, we take the output mask of baseline models and then create a contour around it. This contour can be convex or not. We then apply a convex hull on this contour to turn it into a convex polygon. Lastly, we fill in this new contour to create a new mask in a desired convex shape. This procedure is demonstrated in Figure \ref{fig:image2}.

Furthermore, it is worth noting that many algorithms have been derived to create a convex hull around a set of points. Works on this can be found in \cite{barber1996quickhull,eddy1977new,chazelle1993optimal}. These can all be applied to our algorithm at the step of creating the convex hull around the contour.

\section{Experiments}

In this section, we will present all necessary experiments related to the proposed technique. We consider different scenarios and compare three different learning models for the problem of kidney pathology segmentation with noise and normal data. 

\subsection{Datasets}

\begin{figure}[ht]
    \centering
    \begin{subfigure}[b]{0.3\textwidth}
        \includegraphics[width=\linewidth]{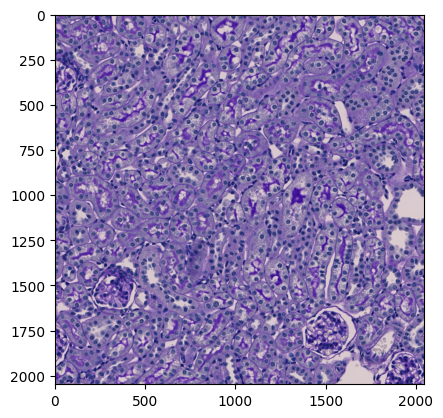}
    \end{subfigure}
    \hfill
    \begin{subfigure}[b]{0.3\textwidth}
        \includegraphics[width=\linewidth]{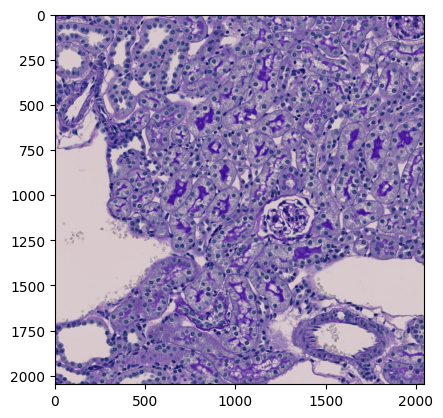}
    \end{subfigure}
    \hfill
    \begin{subfigure}[b]{0.3\textwidth}
        \includegraphics[width=\linewidth]{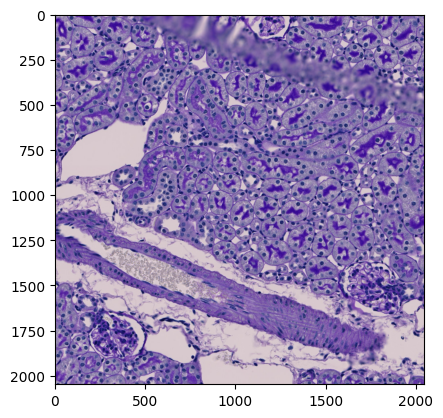}
    \end{subfigure}
    
    \vspace{0.5cm}
    
    \begin{subfigure}[b]{0.3\textwidth}
        \includegraphics[width=\linewidth]{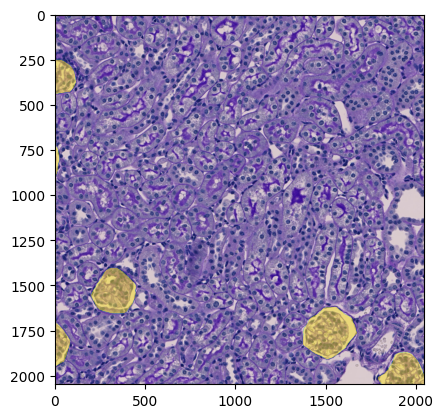}
    \end{subfigure}
    \hfill
    \begin{subfigure}[b]{0.3\textwidth}
        \includegraphics[width=\linewidth]{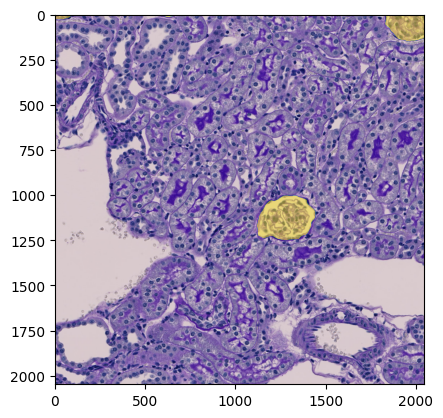}
    \end{subfigure}
    \hfill
    \begin{subfigure}[b]{0.3\textwidth}
        \includegraphics[width=\linewidth]{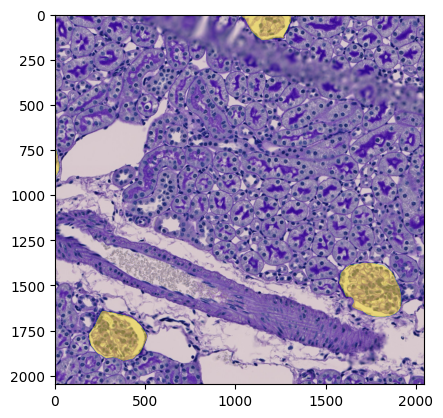}
    \end{subfigure}

    \caption{Examples of Dataset}
    \label{fig:2x3grid}
\end{figure}

Related to the dataset, we experiment on the dataset of Kidney Pathology images from KPIs2024 AI Challenge\footnote{\url{https://sites.google.com/view/kpis2024}} in this work. 
The dataset consists of whole slide images (WSIs) obtained from four different groups of mouse models:
\begin{enumerate}[label=(\roman*)]
\item Normal group: normal mice sacrificed at the age of 8 weeks.
\item 5/6Nx group: mice underwent 5/6 nephrectomy, sacrificed at 12 weeks after nephrectomy (age of 20 weeks). 
\item DN group: eNOS-/-/ lepr(db/db) double-knockout mice, sacrificed at the age of 18 weeks.
\item NEP25 group: transgenic mice that express human CD25 selectively in podocytes (NEP25), sacrificed at 3 weeks after immunotoxin-induced glomerular injury (age of 11 weeks).
\end{enumerate}
The tissue sections were stained using Periodic acid Schiff (PAS) to highlight cellular and structural elements. Each image features nephrons containing a glomerulus and a small cluster of blood vessels. The stained slides were digitized at Vanderbilt University Medical Center. The digital images have been annotated by experienced pathologists.\\
More precisely, the training dataset consists of 30 whole slide images (WSIs) and includes 5,331 high-resolution patches (2048 × 2048) extracted from them. Meanwhile, the test dataset contains 12 whole slide images with a total of 2,305 high-resolution patches (2048 × 2048) taken from these slides.\\
Figure \ref{fig:2x3grid} illustrates three examples of the data set. The first row indicates the original data, while the second contains a mask around the glomerulus.

\subsection{Experimental Design}
To evaluate the effectiveness of the CovHuSeg post-processing method when the amount of training data is limited, we trained the selected models on four separate splits of the kidney pathology image segmentation (KPIs) challenge dataset:
\begin{enumerate}[label=(\roman*)]
    \item Split A: 50\% randomly selected subjects for each group of mouse models, with 100\% of patch images per subject;
    \item Split B: 100\% subjects for each group of mouse models, with 50\% randomly selected patch images per subject;
    \item Split C: 50\% randomly selected subjects for each group of mouse models, with 50\% randomly selected patch images per subject;
    \item Split D: 100\% subjects for each group of mouse models, with 25\% randomly selected patch images per subject.
\end{enumerate}

We trained four different models for the main problem, including UNet \cite{ronneberger2015u}, UNet++ \cite{zhou2018unet++}, UNet3+ \cite{huang2020unet}, and TransUNet \cite{chen2024transunet}, on each of these four splits for 150 epochs with weighted Dice Loss and Binary Cross Entropy Loss ($0.7 * \mathcal{L}_\text{dice} + 0.3 * \mathcal{L}_\text{BCE}$) using Adam \cite{kingma2017adammethodstochasticoptimization} optimizer with a learning rate of $0.001$.

To further investigate how the CovHuSeg post-processing method can boost the inference's performance on noisy cases, we generated an additional test dataset by adding Gaussian noise ($\text{std}=0.28$) to the original test dataset.

\subsection{Results}
Overall, the performance of UNet \cite{ronneberger2015u} is the highest among the models investigated, while TransUnet \cite{chen2024transunet} has the lowest performance on both normal and noisy test datasets. All models suffer from artificial Gaussian noise in the test images, as can be seen by a noticeable drop in the Dice Score of the noisy test dataset. However, our proposed CovHuSeg method is more effective for models that are originally poorly performing models. The absolute increase in the Dice Score metric ranges from $0.005$ to $0.032$ for the normal test dataset and $0.008$ to $0.033$ for the noisy test dataset. The detailed experiment results are reported in Table \ref{tab:results_normal} and Table \ref{tab:results_noise}.

One can see that applying the CovHuSeg technique can improve the performance of these models compared to using the proposed method for both two cases: normal testing dataset and noisy testing dataset. The relative percentage of improvements in terms of the Dice scores varies from 0.55\% to 5.64\% when compared to the normal testing dataset. Meanwhile, using the CovHuSeg method can help to gain a better increase on the Dice scores on the noisy testing dataset. 
\begin{table}[!ht]
\centering
\caption{Dice scores, together with an absolute and relative increase for an experiment on the normal testing dataset.}
\vspace{5mm}
\begin{tabular}{|c|c|c|c|c|c|}
\hline
\textbf{Model} & \textbf{Split} & \textbf{Without CovHuSeg} & \textbf{With CovHuSeg} & \textbf{Increase} & \textbf{$\uparrow\%$} \\
\hline
UNet \cite{ronneberger2015u} & A & 0.835 & 0.841 & 0.006 & 0.74\% \\
UNet \cite{ronneberger2015u} & B & 0.858 & 0.863 & 0.005 & 0.55\% \\
UNet \cite{ronneberger2015u} & C & 0.843 & 0.850 & 0.007 & 0.78\% \\
UNet \cite{ronneberger2015u} & D & 0.848 & 0.853 & 0.005 & 0.60\% \\
\hline
UNet++ \cite{zhou2018unet++} & A & 0.837 & 0.842 & 0.006 & 0.67\% \\
UNet++ \cite{zhou2018unet++} & B & 0.851 & 0.859 & 0.008 & 0.90\% \\
UNet++ \cite{zhou2018unet++} & C & 0.837 & 0.844 & 0.007 & 0.79\% \\
UNet++ \cite{zhou2018unet++} & D & 0.853 & 0.858 & 0.005 & 0.62\% \\
\hline
UNet3+ \cite{huang2020unet} & A & 0.708 & 0.720 & 0.012 & 1.69\% \\
UNet3+ \cite{huang2020unet} & B & 0.787 & 0.800 & 0.013 & 1.66\% \\
UNet3+ \cite{huang2020unet} & C & 0.715 & 0.733 & 0.018 & 2.49\% \\
UNet3+ \cite{huang2020unet} & D & 0.789 & 0.797 & 0.008 & 1.05\% \\
\hline
TransUnet \cite{chen2024transunet} & A & 0.550 & 0.581 & 0.032 & 5.74\% \\
TransUnet \cite{chen2024transunet} & B & 0.687 & 0.711 & 0.023 & 3.42\% \\
TransUnet \cite{chen2024transunet} & C & 0.569 & 0.595 & 0.026 & 4.56\% \\
TransUnet \cite{chen2024transunet} & D & 0.682 & 0.702 & 0.020 & 2.89\% \\
\hline
\end{tabular}
\label{tab:results_normal}
\end{table}

\begin{table}[!ht]
\centering
\caption{Dice scores, together with an absolute and relative increase for an experiment on the noisy testing dataset.}
\vspace{5mm}
\begin{tabular}{|c|c|c|c|c|c|}
\hline
\textbf{Model} & \textbf{Split} & \textbf{Without CovHuSeg} & \textbf{With CovHuSeg} & \textbf{Increase} & \textbf{$\uparrow\%$} \\
\hline
UNet \cite{ronneberger2015u} & A & 0.800 & 0.809 & 0.010 & 1.20\% \\
UNet \cite{ronneberger2015u} & B & 0.821 & 0.828 & 0.008 & 0.93\% \\
UNet \cite{ronneberger2015u} & C & 0.807 & 0.818 & 0.011 & 1.35\% \\
UNet \cite{ronneberger2015u} & D & 0.815 & 0.822 & 0.008 & 0.97\% \\
\hline
UNet++ \cite{zhou2018unet++} & A & 0.797 & 0.807 & 0.009 & 1.15\% \\
UNet++ \cite{zhou2018unet++} & B & 0.812 & 0.823 & 0.012 & 1.46\% \\
UNet++ \cite{zhou2018unet++} & C & 0.801 & 0.812 & 0.011 & 1.32\% \\
UNet++ \cite{zhou2018unet++} & D & 0.825 & 0.834 & 0.009 & 1.06\% \\
\hline
UNet3+ \cite{huang2020unet} & A & 0.579 & 0.593 & 0.014 & 2.42\% \\
UNet3+ \cite{huang2020unet} & B & 0.697 & 0.717 & 0.020 & 2.87\% \\
UNet3+ \cite{huang2020unet} & C & 0.571 & 0.595 & 0.024 & 4.29\% \\
UNet3+ \cite{huang2020unet} & D & 0.696 & 0.711 & 0.016 & 2.23\% \\
\hline
TransUnet \cite{chen2024transunet} & A & 0.260 & 0.287 & 0.027 & 10.34\% \\
TransUnet \cite{chen2024transunet} & B & 0.481 & 0.514 & 0.033 & 6.76\% \\
TransUnet \cite{chen2024transunet} & C & 0.287 & 0.317 & 0.030 & 10.40\% \\
TransUnet \cite{chen2024transunet} & D & 0.550 & 0.581 & 0.031 & 5.56\% \\
\hline
\end{tabular}
\label{tab:results_noise}
\end{table}

\section{Conclusion and Future Works}
In this paper, we have proposed and examined the post-processing method CovHuSeg and implemented a variety of experiments to show its effectiveness in improving the accuracy of baseline models for the segmentation task on the Kidney Pathology images dataset. In particular, we experimented with the UNet \cite{ronneberger2015u}, UNet++ \cite{zhou2018unet++}, UNet3+ \cite{huang2020unet}, and TransUNet \cite{chen2024transunet} models. For each experiment, we also tested different values of training subjects and training samples per subject. Furthermore,  in some experiments, noises are added to the dataset to test the adaptability of our proposed method to noisy environments. \\
Overall, the CovHuSeg algorithm helps increase the accuracy of all models, whether the data contains noise or is noiseless. Additionally, our method performs much better in scenarios with limited data.

The new CovHuSeg post-processing will lay the foundation for many interesting future works. For example, thresholding methods can be studied and applied simultaneously with the CovHuSeg algorithm to further improve the accuracy of the outcome convex masks. Another approach for future improvements would be to incorporate the CovHuSeg algorithm into the training process of segmentation models. This would require a loss function that, after being optimized, would guarantee the outcome mask to be convex.

\section*{Acknowledgment}
We would like to thank AISIA Lab and the University of Science, Vietnam National University Ho Chi Minh City, for supporting us during this project.

%
%
\bibliographystyle{splncs04}
\bibliography{mybibliography}

\end{document}